\documentclass{article}
\usepackage{amsfonts,amssymb, amsmath}
\usepackage[english]{babel}
\usepackage{graphicx}

\def\nn{\nonumber }
\def\bq{ \begin{equation}}
\def\eq{ \end{equation}}
\def\ben{ \begin{eqnarray}}
\def\en{ \end{eqnarray}}
\def\g{{\gamma}}

\newtheorem{prop}{Proposition}

\newtheorem{re}{Remark}
\newenvironment{rem}{\begin{re} \rm }{\end{re}}

\newtheorem{exa}{Example}

\begin{document}

\title{B\"{a}cklund transformations for the nonholonomic Veselova system}
\author{A.V. Tsiganov \\
\it\small St.Petersburg State University, St.Petersburg, Russia\\
\it\small e-mail:  andrey.tsiganov@gmail.com}

\date{}
\maketitle

\begin{abstract}
 We present auto and hetero B\"acklund transformations of the  nonholonomic Veselova system using standard  divisor arithmetic on the hyperelliptic curve of genus two. As a  by-product one gets two natural integrable systems on the cotangent bundle to the unit two-dimensional sphere whose additional integrals of motion are  polynomials in the momenta of fourth order.
 \end{abstract}

\section{Introduction}
\setcounter{equation}{0}
In nonholonomic mechanics a special attention is given to the systems whose equations of motion after suitable reduction yield conformally Hamiltonian vector field, which can be studied by the standard methods of Hamiltonian mechanics after an appropriate time reparameterization, see reviews \cite{biz13,bm05,bbm11,bm14,fed04} and the references therein. For instance,  because the generic level set of the integrals of motion is independent of time, we can study symmetries  of this manifold simultaneously for Hamiltonian and non-Hamiltonian vector fields associated with this level set.

The main our aim is to obtain B\"{a}cklund transformations (BT) for the nonholonomic Veselova system \cite{ves86}.  Historically, the B\"{a}cklund transformations originate in differential geometry in the 1880s and, in particular, they arose as certain transformations between surfaces.  In the theory of integrable systems, according to classical definition by Darboux \cite{darb}, a B\"{a}cklund transformation (BT) between the two given PDEs
\[H(u,x,t)=0\quad\mbox{and}\quad \tilde{H}(\tilde{u},\tilde{x},\tau)=0\]
is a pair of relations
\bq \label{ff-bk}
F_{1,2}(u,x,t,\tilde{u},\tilde{x},\tau)=0
\eq
and some additional relations between $(x, t)$ and $(\tilde{x}, \tau)$, which allow to get both equations  $H$ and $\tilde{H}$, see details in \cite{ab81,rod02}. The BT is called an auto-BT or a hetero-BT depending whether the two PDEs are the same or not.  Auto BTs are used to generate solutions of the given partial differential equation, starting from known solutions, even trivial ones.  The hetero BTs describe a correspondence between two different PDE's rather than a one-to-one mapping between their solutions.

In classical mechanics, auto BT of the Hamilton-Jacobi partial differential equation $H=E$  is a canonical transformation of variables
  \bq\label{btrans}
    \mathcal B:\qquad (u,p_u)\to (\tilde{u},\tilde{p}_u)
  \eq
on the phase space $\mathcal M$ preserving the form of  this Hamilton-Jacobi equation  \cite{toda75,stef82}. There are also other definitions associated with Lax pair \cite{skl00} or  with the corresponding algebraic curve \cite{kuz02},  but we prefer to use definition related to the Hamilton-Jacobi equation itself. A comprehensive definition of hetero BT is also not yet available. Some examples the hetero BTs relating different Hamilton-Jacobi equations were considered in \cite{ts15d, ts15a,ts15b,ts15c}.

For many integrable by quadratures dynamical systems in holonomic and nonholonomic mechanics one faces the Abel quadratures, which for the Veselova system  with were obtained in original paper \cite{ves86}.  In these cases solutions of equations of motion are given by Jacobi inversion of Abel quadratures and the corresponding Abel map allows us to  relate the generic level set of integrals of motion with the  Jacobian $Jac(\mathcal C)$ of some algebraic curve $\mathcal C$, which has a well-studied group structure.   It also allows us to consider original variables $ (u, p_u) $   and their images $(\tilde{u},\tilde{p}_u)$ on $\mathcal M$ as coordinates of two reduced divisors $D$ and $\tilde{D}$ on $Jac(\mathcal C)$ and to identify any auto BT  (\ref{btrans}) with a composition of suitable group operations
\begin{equation}\label{add-jac}
D\approx \tilde{D},\quad  D+ D'=\tilde{D}\qquad\mbox{and}\qquad [\ell] D=\tilde{D}\,,
\end{equation}
where $\approx$, $+$ and $[\ell]$ denote equivalence, addition and scalar multiplication by an integer and
$D'$ is an auxiliary divisor depending on arbitrary parameters.

In \cite{cant87} Cantor proposed a concrete algorithm for performing computations in Jacobian groups of hyperelliptic curves which consists of two stages: the composition stage, which generally outputs an unreduced divisor, and the reduction stage, which transforms the unreduced divisor into the unique reduced divisor.  Now we have a lot of algorithms and their professional computer implementations for the divisor arithmetics on low, mid and high-genus curves, for generic divisor doubling, tripling etc, see references in \cite{cost12,hand06,l05}.

In Hamiltonian mechanics, authors usually consider only full degree divisors $D, \tilde{D} $  and weight one divisor  $D'$ in (\ref{add-jac}) that makes  the reduction stage  of Cantor's algorithm trivial. This partial group operation and the corresponding  auto BTs  have  been  studied from the different points of view in many publications, in particular see \cite{fed00,kuz02,skl00} and references therein. In fact, this very special  construction of auto BTs  was developed for  simplest one-parametric discretization of original continuous systems.

 Below we discuss auto BTs associated with addition of two generic divisors and with generic divisor doubling when reduction coincides with inversion. These auto BTs represent hidden symmetries of the level manifold  which yields new canonical variables on original phase space and, therefore, we can use these new variables to  to construction of new integrable systems, i.e. to construction of hetero BTs. In our opinion it is natural to use different types of auto BTs for the different purposes.

\subsection{Veselova system}

Following \cite{ves86}, consider  the motion of the rigid body with a fixed point under nonholonomic
constraint
\bq\label{rel-ves}
(\Omega,\gamma)=0\,.
\eq
Here $\Omega=(\Omega_1,\Omega_2,\Omega_3)$ is the vector of the angular velocity in the body frame, $\gamma=(\g_1,\g_2,\g_3)$ is a unit vector which is fixed in a space frame,  $(x,y)$ and $x \times y$ denote the scalar and  vector products  in $\mathbb R^3$, respectively. Isomorphism of this system with the system describing the motion of a nonhomogeneous ball without slipping and twisting on a plane was found in \cite{bm08}, see also \cite{bmb13}.

A standard form of the equations of motion describing rotation of a rigid body around a fixed point  is the following:
\bq\label{ves-eqm}
 \dfrac{d}{d\tau} {\gamma}=\gamma\times \Omega\,,\qquad \dfrac{d}{d\tau}{M}=M\times \Omega +\lambda \gamma\,.
\eq
The last term in the second equation in (\ref{ves-eqm}) is related to  condition that   projection of the angular velocity $\Omega$ to a fixed vector $\g$  must zero (\ref{rel-ves}).

Here $M=I\Omega$ is the vector of kinetic momentum of the body, expressed in the  body frame.
This frame is firmly attached to the body, its origin is in the body's fixed point, and its axes coincide with the principal inertia axes
of the body. The inertia tensor of the body in this frame is diagonal
\[
 { I}=\left(
 \begin{array}{ccc}
  {I}_1 & 0 & 0 \\
 0 &  {I}_2 & 0 \\
 0 & 0 &  {I}_3 \\
 \end{array}
 \right)\,,\qquad I_1,I_2,I_3>0\,.
\]
Using the equations of motion (\ref{ves-eqm}) and constraint (\ref{rel-ves}) we can calculate  Lagrangian multiplier $\lambda$
 \[
\lambda=\dfrac{( A M\times M,  A \gamma)}{( A \gamma,\gamma)}\,,\qquad A=I^{-1}=\left(
 \begin{array}{ccc}
  {a}_1 & 0 & 0 \\
 0 &  {a}_2 & 0 \\
 0 & 0 &  {a}_3 \\
 \end{array}
 \right)
\]
 Equations of motion  (\ref{ves-eqm}) determine vector field $Z$ on the  manifold $\mathcal M=\mathbb R^3\times so^*(3)$ with coordinates  $z=(\g_1,\g_2,\g_3,M_1,M_2,M_3)$:
 \bq\label{z-ves}
 \dfrac{d}{d\tau} z_i=Z_i\,,\qquad i=1,\cdots,6.
 \eq
 Vector field $Z$  possesses four independent first integrals
\bq\label{int-ves}
 {H}_1=(M, \Omega)\,,\qquad  {H}_2=(M,M)-(\gamma,M)^2\,,\qquad  {C}_1=(\gamma,\gamma)=1,\qquad
 {C}_2=(\gamma,\Omega)=0\,,
\eq
and an invariant measure
\bq\label{rho-ves}
\mu=  {g}\,\mathrm d \gamma \mathrm dM\,,\qquad  {g}={ ( \gamma, A \gamma)}^{1/2}=\sqrt{a_1\g_1^2+a_2\g_2^2+a_3\g_3^2}\,.
\eq
Hence,  it is integrable by quadratures according  the Euler-Jacobi theorem \cite{ves86}.

In order to get these quadratures  we introduce the following Poisson bivector on $\mathcal M$
 \bq\label{poi-ves}
P=\left(\begin{smallmatrix}
        0 &  0 &  0 &\frac{\g_1\g_2\g_3(a_3-a_2)}{g}  &  \g_3\left(g+\frac{(a_3-a_2)x_2^2}{g}\right)  &  -\g_2\left(g-\frac{(a_3-a_2)x_3^2}{g}\right) \\
       0 &  0 &   0 &  -\g_3\left(g-\frac{(a_1-a_3)x_1^2}{g}\right)  &\frac{\g_1\g_2\g_3(a_1-a_3)}{g}   &   \g_1\left(g+\frac{(a_1-a_3)x_3^2}{g}\right)\\
       0 &  0 &  0 &  \g_2\left(g-\frac{(a_1-a_2)x_1^2}{g}\right) & -\g_1\left(g+\frac{(a_1-a_2)x_2^2}{g}\right)  &\frac{\g_1\g_2\g_3(a_2-a_1)}{g} \\
     *  & * & * &   0 & \frac{b_3}{g} &-\frac{b_2}{g}  \\
       * & *  & * &-\frac{b_3}{g} &   0 &  \frac{b_1}{g}\\
       * & * &*  &\frac{b_2}{g}  & -\frac{b_1}{g} &   0 \\
    \end{smallmatrix}\right)
  \,,
\eq
where vector  $b=(b_1,b_2,b_3)$ is equal to
\[
b=(\g,\g)AM-(A\g\times \g)\times M= \left(
                                    \begin{smallmatrix}
                                    (\g,\g)a_1M_1-\g_1\bigl((a_2-a_1)\g_2M_2+(a_3-a_1)\g_3M_3\bigr)\\ \\
                                     (\g,\g)a_2M_2-\g_2\bigl((a_1-a_2)\g_1M_1+(a_3-a_2)\g_3M_3\bigr)  \\ \\
                                     (\g,\g)a_3M_3-\g_3\bigl((a_1-a_3)\g_1M_1+(a_2-a_3)\g_2M_2\bigr)  \\
                                    \end{smallmatrix}
                                  \right)\,.
\]
This bivector $P$ may be obtained using Chaplygin method of reducing multiplier \cite{bm05,fed04},  Poisson reduction on Lie groups \cite{nar07} or the Turiel deformations of the canonical Poisson structures \cite{ts12}.

 It is easy to prove that vector field $Z$ (\ref{z-ves}) is a conformally Hamiltonian vector field on the phase space $\mathcal M=T^*\mathbb S^2$
\[
Z=\dfrac{1}{2g}\,X\,,\qquad X=PdH_1\,,
\]
where Hamiltonain $H_1$ is given by (\ref{int-ves}).

In the next Section we present auto and hetero B\"{a}cklund transformations for these vector fields  $Z$ and $X$ having a common level set of integrals of motion.

\section{Abel equations and auto B\"{a}cklund transformations}
\setcounter{equation}{0}
According \cite{ves86} we can integrate  original equations of motion (\ref{ves-eqm}) after change of  time $d\tau\to 2g dt $, i.e. after transition from conformally Hamiltonian vector field  $Z$ to Hamiltonian vector field $X$. Indeed, let us introduce variables
$u_{1,2}$ and $p_{u_{1,2}}$ using equations
 \bq\label{ves-gamma}
\g_i=\sqrt{\dfrac{(u_1-I_i)(u_2-I_i)}{(I_j-I_i)(I_k-I_i)}},\qquad i\neq j\neq k
\eq
and
\bq\label{ves-M}
M_i=\dfrac{2}{ gI_jI_k}\cdot\dfrac{\varepsilon_{ijk}\g_j\g_k(I_j-I_k)}{u_1-u_2}\Bigl((I_i-u_1)p_{u_1}-(I_i-u_2)p_{u_2}\Bigr)\,,
\eq
where $\varepsilon_{ijk}$ is a totally skew-symmetric tensor. In this variables Poisson bivector (\ref{poi-ves}) has the following   form
\[
P=\left(
    \begin{array}{cccc}
      0 & 0 & 1 & 0 \\
      0 & 0 & 0 & 1 \\
      -1 & 0 & 0 & 0 \\
      0 & -1 & 0 & 0 \\
    \end{array}
  \right)\,,
\]
i.e.  variables $u_{1,2}$ and $p_{u_{1,2}}$ are  canonical  variables
\bq\label{poi1}
\{u_i,p_{u_j}\}=\delta_{i,j}\,,\qquad \{u_1,u_2\}=\{p_{u_1},p_{u_2}\}=0\,.
\eq
We can identify $u_{1,2}$  with the  standard elliptic or spheroconical  coordinates on the unit sphere
\bq\label{ves-def-u}
\dfrac{\g_1^2}{x-I_1}+\dfrac{\g_2^2}{x-I_2}+\dfrac{\g_3^2}{x-I_3}=\dfrac{(x-u_1)(x-u_2)}{(x-I_1)(x-I_2)(x-I_3)}\,,\qquad I_k=\dfrac{1}{a_k}
\eq
The defining equation (\ref{ves-def-u})  should be interpreted as an identity with respect to $x$, and for each set of elliptic coordinates $u_{1,2}$ it is possible to solve (\ref{ves-def-u}) for $\g_i$ by calculating the residues at $x = I_i$.  Notice also that (\ref{ves-def-u}) implies $(\g,\g)=1$ and   $I_1<I_2<I_3$, so that the elliptic coordinate system is orthogonal, and the coordinates $u_{1,2}$ take values
only in the intervals
\[I_1 < u_1 < I_2 < u_2 <  I_3.\]
In geometry using a simultaneous rescaling of the coordinates and the parameters, $u_i\to au_i+b$ and $I_i \to aI_i+b$, it is always possible to take $I_1=0$ and $I_3=1$. In mechanics $I_k>0$ are the some fixed momenta of inertia which can not be changed.

The corresponding momenta
\bq\label{ves-def-pu}
p_{u_{1,2}}=\dfrac{1}{2g}\left(\dfrac{\g_1(\g_2M_3-\g_3M_2)}{x-I_1}+\dfrac{\g_2(\g_3M_1-\g_1M_3)}{x-I_2}
+\dfrac{\g_3(\g_1M_2-\g_2M_1)}{x-I_3}
\right)_{x=u_{1,2}}
\eq
differ on standard expressions of momenta via angular momentum by the factor $g=\sqrt{(\g,A\g)}$ in the denominator. It would do no harm  us to identify phase space $\mathcal M$ with  the cotangent bundle $T^*\mathbb S^2$  of the unit sphere \cite{bm05,fed04,ves86}.

In this canonical variables $T^*\mathbb S^2$ original Hamiltonians (\ref{int-ves}) are equal to
\bq\label{ves-ham-u}
H_1=\dfrac{ \varphi(u_1)p_{u_1}^2}{u_1-u_2}+\dfrac{\varphi(u_2)p_{u_2}^2}{u_2-u_1}\qquad\mbox{and}\qquad
H_2=\dfrac{ u_2\varphi(u_1)p_{u_1}^2}{u_1-u_2}+\dfrac{u_1\varphi(u_2)p_{u_2}^2}{u_2-u_1}\,,
\eq
where
\bq\label{ves-phi}
\varphi(u)=4u(1-a_1u)(1-a_2u)(1-a_3u)\,.
\eq
To describe evolution of $u_ {1,2} $ with respect to $H_ {1,2} $ we use the canonical Poisson bracket
(\ref{poi1}) and expressions for $H_{1,2}$ to obtain
\begin{equation}\label{ves-eq1}
\frac{du_1}{dt_1}=\{u_1,H_1\}=\frac{2\varphi(u_1) p_{u_1}}{u_1-u_2}\,,\qquad
\frac{du_2}{dt_1}=\{u_2,H_1\}=\frac{2\varphi(u_2) p_{u_2}}{u_2-u_1}
\end{equation}
and
\begin{equation}\label{ves-eq2}
\frac{du_1}{dt_2}=\{u_1,H_2\}=\frac{2u_2\varphi(u_1)p_{u_1}}{u_1-u_2}\,,\qquad
\frac{du_2}{dt_2}=\{u_2,H_2\}=\frac{2u_1\varphi(u_2)p_{u_2}}{u_2-u_1}\,.
\end{equation}
To solve the Hamilton-Jacobi equations $H_{1,2}=h_{1,2}$ with respect to $p_{u_{1,2}}$
\[p_{u_k}^2=\varphi(u_k)^{-1}(h_1u_k-h_2)\,,\qquad k=1,2.\]
and to substitute these expressions  into (\ref{ves-eq1}-\ref{ves-eq2} )  one gets standard Abel quadratures
\bq\label{ves-ab-q1}
\frac{du_1}{\sqrt{f(u_1)}}+\frac{du_2}{\sqrt{f(u_2)}}=2dt_2\,,\qquad \frac{u_1du_1}{\sqrt{f(u_1)}}+\frac{u_2du_2}{\sqrt{f(u_2)}}=2dt_1,
\eq
on the  hyperelliptic curve $\mathcal C$ of genus two  defined by equation
\bq\label{ves-eq-c}
\mathcal C:\quad y^2=f(x)\,,\qquad f(x)=4x(a_1x-1)(a_2x-1)(a_3x-1)(h_1x-h_2)\,.
\eq
Thus,   we can identify the generic level set of the first integrals for the Veselova system with a symmetric product $\mathcal C\times \mathcal C$ and later with Jacobi variety $Jac(\mathcal C)$ of hyperelliptic curve $\mathcal C$ of genus two  \cite{ves86}.  It is a starting point in the construction of the auto and hetero B\"{a}cklund transformations of the Veselova system.

\begin{rem}
Multidimensional generalizations of the Veselova problem of a nonholonomic rigid body motion and  Abel quadratures for the corresponding Hamiltonian vector fields are discussed in \cite{fed04}. We can directly apply these quadratures to construction of the B\"{a}cklund transformations in the framework of the Abel theory \cite{ts16m}.
\end{rem}

\subsection{Abel equations and group operations}
Suppose that transformation of variables
\bq\label{st-b-ves}\mathcal B:\quad (u_{1},u_2,p_{u_1},p_{u_2})\to
(\tilde{u}_{1},\tilde{u}_2,\tilde{p}_{u_1},\tilde{p}_{u_2})
\eq
preserves Hamilton equations (\ref{ves-eq1}-\ref{ves-eq2}) and the form of Hamiltonians (\ref{ves-ham-u}). It means that new variables satisfy to algebraic equations
\[
\begin{array}{c}
\dfrac{ \varphi(u_1)p_{u_1}^2}{u_1-u_2}+\dfrac{\varphi(u_2)p_{u_2}^2}{u_2-u_1}=H_1=
\dfrac{ \varphi(\tilde{u}_1)\tilde{p}_{u_1}^2}{\tilde{u}_1-\tilde{u}_2}+\dfrac{\varphi(\tilde{u}_2)\tilde{p}_{u_2}^2}{\tilde{u}_2-\tilde{u}_1}
\\
\\
\dfrac{ u_2\varphi(u_1)p_{u_1}^2}{u_1-u_2}+\dfrac{u_1\varphi(u_2)p_{u_2}^2}{u_2-u_1}=H_2=\dfrac{ \tilde{u}_2\varphi(\tilde{u}_1)\tilde{p}_{{u}_1}^2}{\tilde{u}_1-\tilde{u}_2}+\dfrac{\tilde{u}_1\varphi(\tilde{u}_2)\tilde{p}_{{u}_2}^2}{\tilde{u}_2-\tilde{u}_1}\,,
\end{array}
\]
and differential equations
\bq\label{ves-ab-q2}
\frac{d\tilde{u}_1}{\sqrt{f(\tilde{u}_1)}}+\frac{d\tilde{u}_2}{\sqrt{f(\tilde{u}_2)}}=2dt_2\,,\qquad \frac{\tilde{u}_1d\tilde{u}_1}{\sqrt{f(\tilde{u}_1)}}+\frac{\tilde{u}_2d\tilde{u}_2}{\sqrt{f(\tilde{u}_2)}}=2dt_1\,.
\eq
Subtracting (\ref{ves-ab-q2}) from (\ref{ves-ab-q1}) one gets two Abel differential equations
\begin{equation}\label{ab-eq-g2}
\begin{array}{c}
\omega_1(x_1,y_1)+\omega_1(x_2,y_2)+\omega_1(x_3,y_3)+\omega_1(x_4,y_4)=0\,,\\ \\
\omega_2(x_1,y_1)+\omega_2(x_2,y_2)+\omega_2(x_3,y_3)+\omega_2(x_4,y_4)=0\,,
\end{array}
\end{equation}
where
 \bq\label{xy-ves}
 x_{1,2}=u_{1,2},\quad y_{1,2}=\varphi(u_{1,2})p_{u_{1,2}}\,,\qquad
x_{3,4}=\tilde{u}_{1,2},\quad y_{3,4}=-\varphi(\tilde{u}_{1,2})\tilde{p}_{u_{1,2}}
\eq
and $\omega_{1,2}$ form a base of holomorphic differentials on hyperelliptic curve $\mathcal C$ of genus two
\[
\omega_1(x,y)=\frac{dx}{y}\,,\qquad \omega_2(x,y)=\frac{xdx}{y}\,.
\]
Here we change sign of $y_{3,4}$ by using standard hyperelliptic involution $(x,y)\to (x,-y)$. Historical perspective and modern  geometric meaning of  Abel differential equations are discussed in \cite{gg,kl05}.

\begin{rem}
Abel equations are closely related with the group law in the Jacobian of the corresponding algebraic curve  \cite{cant87, fed00, hand06,kuz02, ts16m}. In fact, there are two main methods for deriving  the group law in Jacobian of  hyperelliptic curve: the algebraic method  \cite{l05} based on Harley's formulation of Cantor's algorithm, and the geometric method using interpolation of points \cite{cost12}, which is based on  Clebsch's geometric formulation of the Abel theorem.
In both cases,  we can  either use an abstract modern language of algebraic geometry, which is somewhat difficult to understand experts in mechanics, physics, practical cryptography etc, either use some trivial geometric construction,  which allow us to  get the necessary explicit formulae.
\end{rem}

 For the low-genus hyperelliptic curves and the corresponding mechanical systems it is easy to explain all the necessary explicit formulae  in the framework of the classical Abel theory, i.e. without the modern abstractions applicable to all the possible cases. Indeed, let us consider  Abel differential equations (\ref{ab-eq-g2}) on the genus two hyperelliptic curve
 \bq\label{eq-c}
 \mathcal C:\qquad y^2=\mathrm a_5x^5+\mathrm a_4x^4+\mathrm a_3x^3+\mathrm a_2x^2+\mathrm a_1x+\mathrm a_0\,.
 \eq
 According to Abel's idea, solutions $(x_i,y_i) $ are abscissas and ordinates of  the points  of  $\mathcal C$ intersecting with the second plane curve defined by equation
\bq\label{eq-p}
y-P(x)=0\,,\qquad P(x)=b_3x^3+b_2x^2+b_1x+b_0\,.
\eq
 In our case four points  of intersection $\mathrm p_1=(x_1,y_1),\ldots\mathrm p_4=(x_4,y_4)$ are solutions of Abel equations (\ref{ab-eq-g2}), whereas two remaining points $\mathrm p_{5,6}$ are arbitrary. Using this freedom we can suppose that (see Figure 1):
\begin{enumerate}
  \item points $\mathrm p_{5,6}$ are given at some fixed finite positions (three full degree divisors);
  \item  one point $\mathrm p_6$ is at infinity (two full degree divisors and divisor weight one);
  \item  points $\mathrm p_{5,6}$ coincides with $\mathrm p_{1,2}$ (generic divisor doubling).
\end{enumerate}
\begin{figure}[!ht]
\begin{minipage}[h]{0.33\linewidth}
\center{\includegraphics[width=0.85\linewidth]{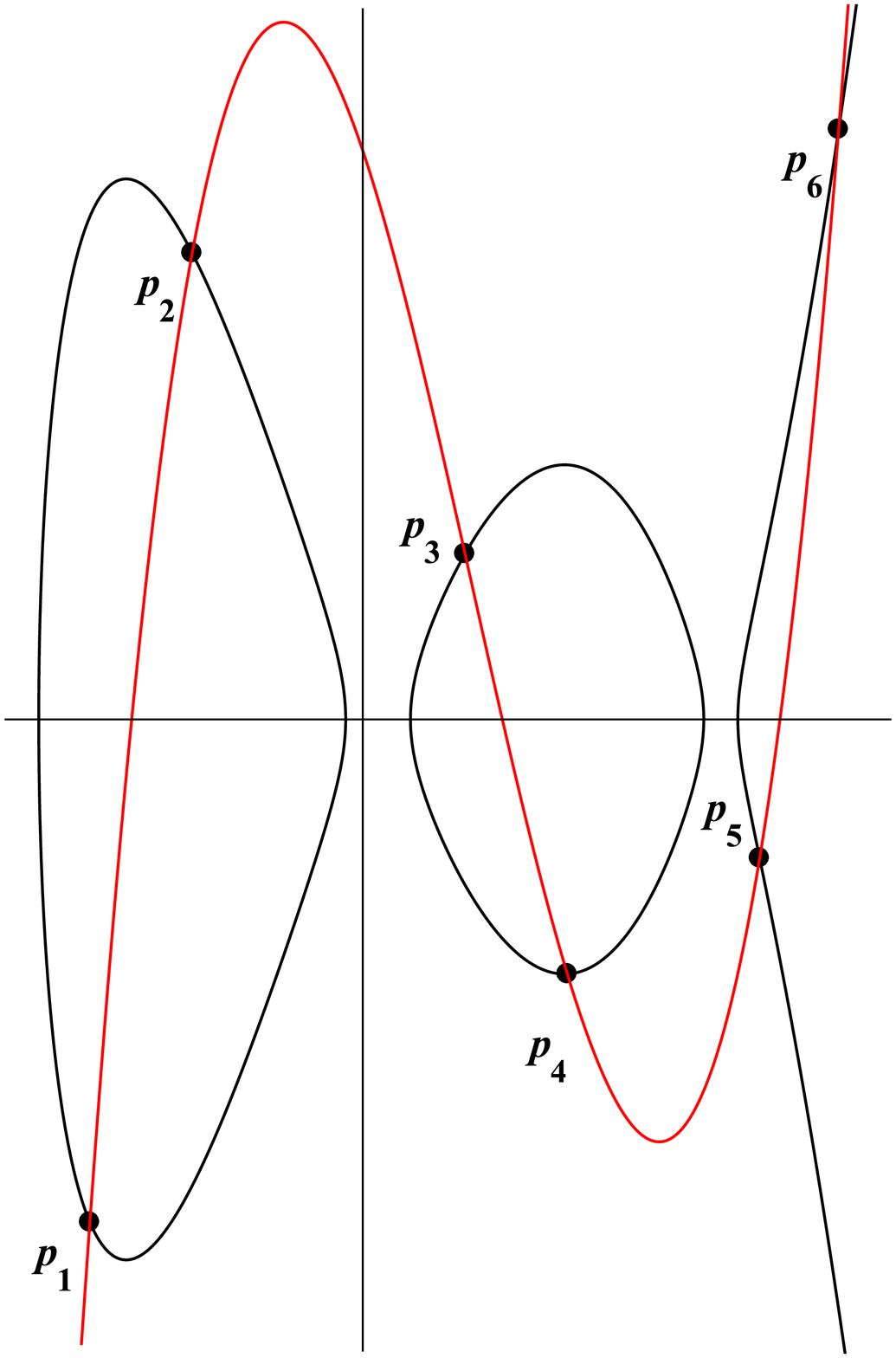} \\ Case 1}
\end{minipage}
\begin{minipage}[h]{0.33\linewidth}
\center{\includegraphics[width=0.85\linewidth]{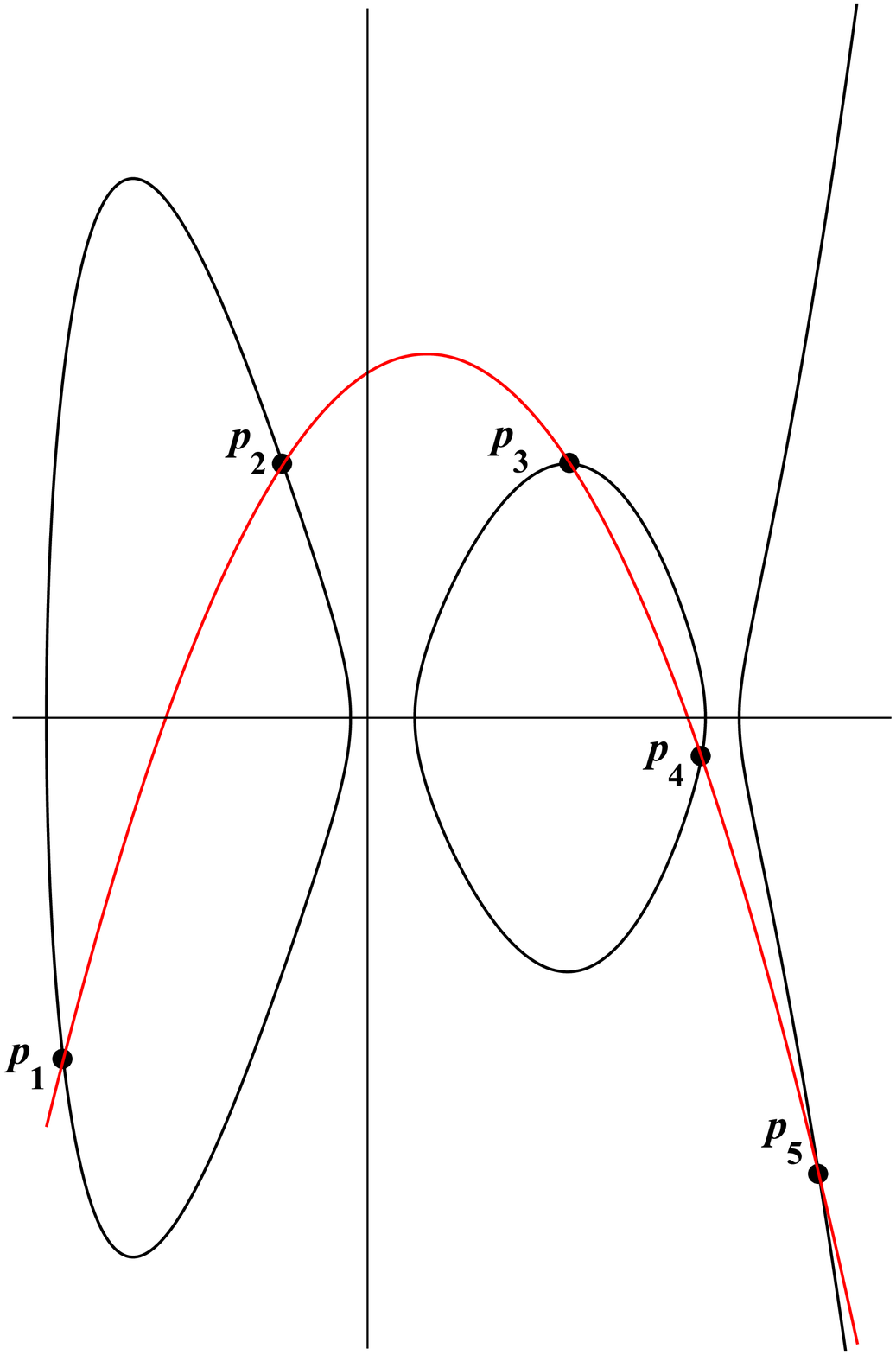} \\ Case 2}
\end{minipage}
\begin{minipage}[h]{0.33\linewidth}
\center{\includegraphics[width=0.85\linewidth]{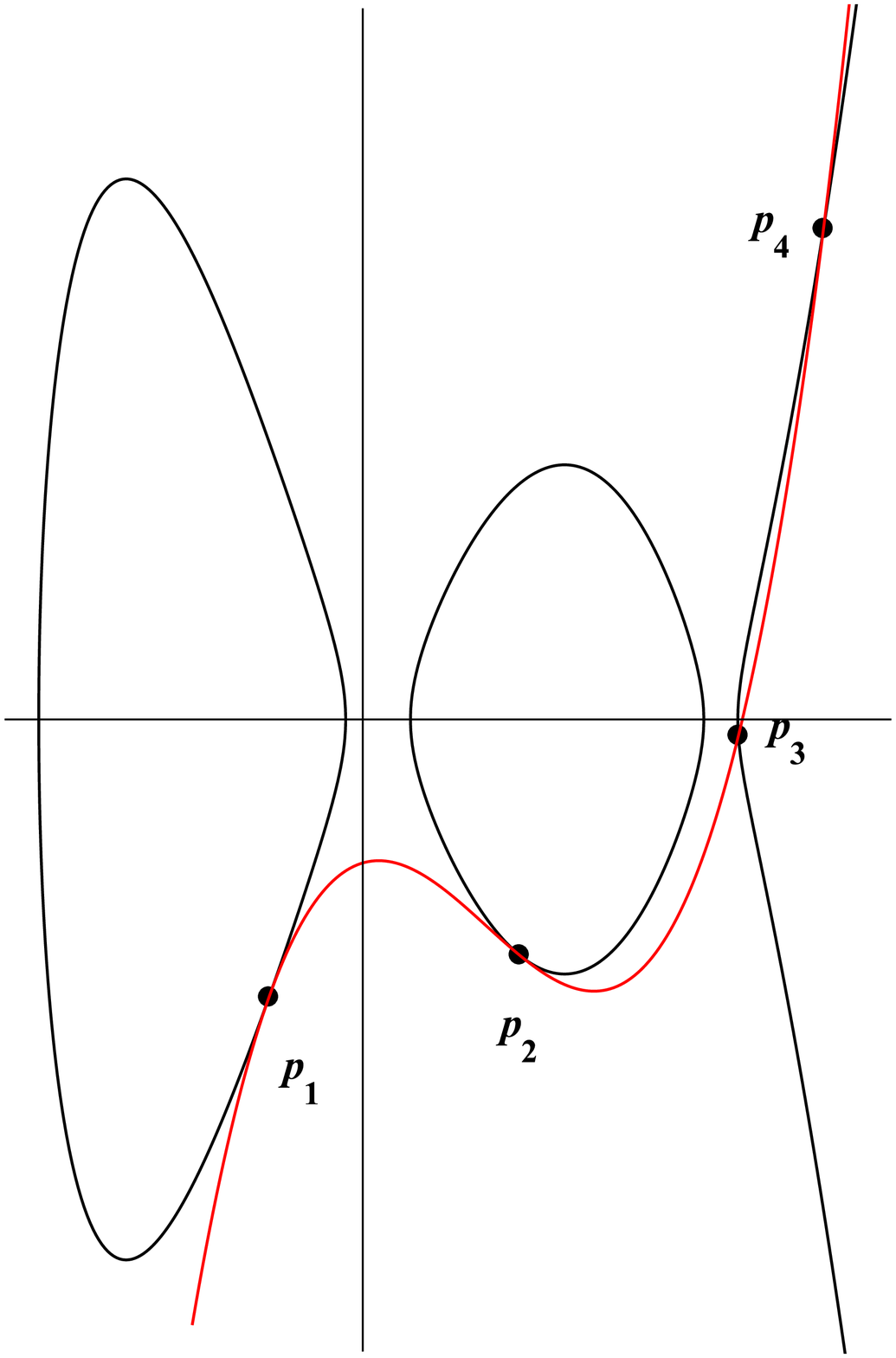} \\ Case 3}
\end{minipage}
\caption{Standard picture from the paper on hyperelliptic curve cryptography \cite{cost12}.}
\end{figure}
\begin{rem}
There are other  reduced divisors  on $Jac(\mathcal C)$ \cite{cant87}. Here we consider only divisors associated with the given  integrable system with two degrees of freedom (\ref{ves-eq1}-\ref{ves-eq2}).  Examples of auto BTs associated with reduced divisors at Case 1 and at  Case 2 may be found in \cite{fed00,skl00,kuz02}.  Examples of auto BTs associated with a generic divisor doubling (Case 3)  may be found in \cite{ts17d}.
\end{rem}

Substituting $y=P(x)$ into the equation of the curve (\ref{eq-c})  we obtain so-called Abel polynomial
 \[
\psi(x)=P(x)^2-f(x)=(b_3x^3+b_2x^2+b_1x+b_0)^2
-(\mathrm a_5x^5+\mathrm a_4x^4+\mathrm a_3x^3+\mathrm a_2x^2+\mathrm a_1x+\mathrm a_0) \,,
\]
which has no multiple roots in the  first and second cases and two double roots in the third case:
\[
\begin{array}{cl}
1.&\quad\psi(x)=b_3^2(x-x_1)(x-x_2)(x-x_3)(x-x_4)(x-x_5)(x-x_6)\,,\\
\\
2.&\quad \psi(x)=-\mathrm a_5(x-x_1)(x-x_2)(x-x_3)(x-x_4)(x-x_5)\,,\quad b_3=0\,.\\
\\
3.&\quad \psi(x)=b_3^2(x-x_1)^2(x-x_2)^2(x-x_3)(x-x_4)\,.
\end{array}
\]
Using Abel polynomial $\psi$ we can easily determine abscissas $x_{3,4}$ as functions on coordinates of other points of intersection:
\[
\begin{array}{cl}
1.\quad& (x-x_3)(x-x_4)=\dfrac{\psi(x)}{b_3^2(x-x_1)(x-x_2)(x-x_5)(x-x_6)}\,;\\
\\
2.\quad& (x-x_3)(x-x_4)=\dfrac{\psi(x)}{\mathrm a_5(x-x_1)(x-x_2)(x-x_5)}\,,\\
\\
3.\quad& (x-x_3)(x-x_4)=\dfrac{\psi(x)}{b_3^2(x-x_1)^2(x-x_2)^2}\,.
\end{array}
\]
Equating coefficients is an efficient way to compute the exact division required above  gives:
\ben\label{g2-x1}
1.&x_3+x_4= \dfrac{\mathrm a_5-2b_2b_3}{b_3^2}-x_1-x_2-x_5-x_6\,,
\qquad x_3x_4=\dfrac{2b_1b_3+b_2^2-\mathrm a_4}{b_3^2}-\\
\nn\\
&\phantom{x_3+x_4=}-(x_1+x_2+x_5+x_6)(x_3+x_4)
-x_1(x_2+x_5+x_6) -x_2(x_5+x_6)-x_5x_6
\nn
\en
and similar for the second and third cases
\bq\label{g2-x23}
\begin{array}{cl}
2.\quad & x_3+x_4=-x_1-x_2-x_5+\dfrac{b_2^2-\mathrm a_4}{\mathrm a_5},\\
\\
& x_3x_4=\dfrac{\mathrm a_3-2b_1b_2}{\mathrm a_5}-(x_1+x_2+x_5)(x_3+x_4)-x_1x_2-x_1x_5-x_2x_5\,;\\
\\
3.\quad &
x_3+x_4 = -2x_1-2x_2+\dfrac{\mathrm a_5-2b_2b_3}{b_3^2}\,,\\
\\
 & x_3x_4=\dfrac{2b_1b_3+b_2^2-\mathrm a_4}{b_3^2}-2(x_1+x_2)(x_3+x_4)-x_1^2-4x_1x_2-x_2^2\,.
\end{array}
\eq
Four coefficients $b_3,b_2,b_1$ and $b_0$ of the polynomial $P(x)$ (\ref{eq-p}) are calculated  by solving  four algebraic equations:
\bq\label{eq-b}
\begin{array}{cl}
1.\quad &y_{1,2}=P(x_{1,2})\,,\qquad y_{5,6}=P(x_{5,6})\,;\\
\\
2.\quad &y_{1,2}=P(x_{1,2})\,,\qquad y_{5}=P(x_{5})\,,\qquad b_3=0\,;\\
\\
3.\quad& y_{1,2}=P(x_{1,2})\,,\qquad \left.\dfrac{d P(x)}{dx}\right|_{x=x_{1,2}}=\left.\dfrac{d \sqrt{f(x)}}{dx}\right|_{x=x_{1,2}}
\equiv \dfrac{1}{2y_{1,2}} f'(x_{1,2})\,,
\end{array}
\eq
where $f'(x)$ is derivative of $f(x)$ by $x$. Substituting  coefficients $b_{k}$ into (\ref{g2-x1}-\ref{g2-x23}) one gets abscissas $x_{3,4}$ as functions on $x_{1,2},y_{1,2}$ and $x_{5,6}, y_{5,6}$. The corresponding ordinates $y_{3,4}$  are equal to
\bq\label{g2-y}
y_{3,4}=P(x_{3,4})\,,
\eq
where polynomial $P(x)$ is  given by
\bq\label{ab-p}
\begin{array}{lcl}
1.\quad P(x)
&=&\dfrac{(x-x_6) (x-x_5) (x-x_2 ) y_1}{(x_1-x_5 ) (x_1-x_6) (x_1-x_2)}
+
\dfrac{(x-x_6) (x-x_5) (x-x_1 ) y_2}{(x_2-x_5 ) (x_2-x_6) (x_1-x_2)}\\
\\
&+&
\dfrac{(x-x_6) (x-x_2 ) (x-x_1 ) y_5}{(x_5-x_1 ) (x_5-x_2 ) (x_6-x_5)}
+
\dfrac{(x-x_5) (x-x_2 ) (x-x_1 ) y_6}{ (x_6-x_1 ) (x_6-x_2 )(x_6-x_5)}\,,\\
\\
2.\quad
P(x)&=&
\dfrac{y_1(x-x_2)(x-\lambda)}{(x_1-x_2)(x_1-x_5)}+\dfrac{y_2(x-x_1)(x-x_5)}{(x_2-x_1)(x_2-x_5)}
+\dfrac{y_5(x-x_1)(x-x_2)}{(x_5-x_1)(x_5-x_2)}\,,\\
\\
3.\quad P(x)&=&
\dfrac{(x-x_2)^2 (2x-3 x_1+x_2) y_1}{(x_2-x_1)^3}
+
\dfrac{(x-x_1)^2 (2x+x_1-3 x_2) y_2}{(x_1-x_2)^3}\\
\\
&+&\dfrac{(x-x_2)^2 (x-x_1) f'(x_1)}{2(x_1-x_2)^2 y_1}+
\dfrac{(x-x_1)^2 (x-x_2) f'(x_2)}{2(x_1-x_2)^2 y_2}\,.
\end{array}
\eq
These simple equations (\ref{g2-x1}-\ref{g2-x23}) and (\ref{g2-y}) determine group operations relating points of intersection
$\mathrm p_{1,2}=(x_{1,2},y_{1,2})$ with  $\mathrm p_{3,4}=(x_{3,4},y_{3,4})$.

\begin{rem}
In generic cases we can take any professional  implementation of Cantor's algorithm \cite{cant87, hand06}  or implementations of various improvements and extensions to Cantor's algorithm for mid-  or high-genus curves with general divisor doubling, tripling etc \cite{cost12,hand06}.
\end{rem}

\subsection{Auto B\"{a}cklund transformations}
We identify variables on the phase space $T^*\mathbb S^2$ with abscissas and ordinates of the points of intersection in  (\ref{xy-ves})
\[
 x_{1,2}=u_{1,2},\quad y_{1,2}=\varphi(u_{1,2})p_{u_{1,2}}\,,\qquad
x_{3,4}=\tilde{u}_{1,2},\quad y_{3,4}=-\varphi(\tilde{u}_{1,2})\tilde{p}_{u_{1,2}}\,.
\]
Following  \cite{fed04,kuz02,skl00} we also denote coordinates of the remaining two points of intersection as follows
\[
\mathrm p_5=(x_5,y_5)\equiv(\lambda_1,\mu_1)\,,\qquad \mathrm p_6=(x_6,y_6)\equiv(\lambda_2,\mu_2)\,.
\]
Abscissas $\lambda_{1,2}$ are arbitrary numerical parameters, whereas $\mu_{1,2}=\sqrt{(f(\lambda_{1,2})}$ are non trivial combinations of the integrals of motion, i.e. nontrivial functions on the phase space.

In order to get explicit expressions for $\tilde{u}_{1,2}$ and $\tilde{p}_{u_{1,2}}$ in terms of  original elliptic coordinates $u_{1,2}$ and momenta $p_{u_{1,2}}$ we have to:
 \begin{itemize}
   \item find  coefficients $\mathrm a_5,\ldots,\mathrm a_0$ from   the separation relations (\ref{ves-eq-c},\ref{eq-c});
   \item calculate coefficients $b_3,\ldots,b_0$ from (\ref{eq-b});
   \item substitute these coefficients into (\ref{g2-x1}-\ref{g2-x23}) and solve the resulting equations with respect to $x_{3,4}=\tilde{u}_{1,2}$;
   \item calculate momenta $\tilde{p}_{u_{1,2}}$ substituting $\tilde{u}_{1,2}$ into (\ref{g2-y})
   \[
y_{3,4}=P(x_{3,4})\quad
\Longrightarrow
\quad\tilde{p}_{u_{1,2}}=-\dfrac{P(\tilde{u}_{1,2})}{\varphi(\tilde{u}_{1,2})}\,.
\]
 \end{itemize}
In order to get canonical transformation $\mathcal B$ (\ref{st-b-ves}) in term of original variables $\g$ and $M$ we have to substitute $\tilde{u}_{1,2}$ and $\tilde{p}_{u_{1,2}}$ into (\ref{ves-gamma}) and (\ref{ves-M}) instead of $u_{1,2}$ and $p_{u_{1,2}}$ and use (\ref{ves-def-pu}-\ref{ves-def-pu}). We can easily obtain the desired explicit formulae for $\tilde{\g}$ and $\tilde{M}$ using  any modern  computer algebra system and, therefore, we  do not show these bulky expressions here.

\begin{rem}
We can avoid computer calculations and obtain the same expressions by hand using  known Lax representations for finite dimensional systems \cite{kuz02,skl00}. However, it is easy to prove that  in fact Darboux transformations of the Lax pairs yield only  implicit expressions  for the images of original variables even for the systems with two and three degrees of freedom.
\end{rem}

Using explicit formulae for the transformations $\mathcal B$ we can prove the following statement.
\begin{prop}
Equations (\ref{g2-x1}-\ref{g2-x23}) and (\ref{g2-y})   determine canonical transformations $\mathcal B$ (\ref{st-b-ves}) on $T^*\mathbb S^2$ of valency one and two for which original Poisson bracket (\ref{poi1}) has the following form in new variables
\[\begin{array}{cl}
1,2.\qquad&\{\tilde{u}_i,\tilde{p}_{u_j}\}=\phantom{2}\delta_{i,j}\,,\qquad \{\tilde{u}_1,\tilde{u}_2\}=\{\tilde{p}_{u_1},\tilde{p}_{u_2}\}=0
\\
\\
3.\quad&
\{\tilde{u}_i,\tilde{p}_{u_j}\}=2\delta_{i,j}\,,\qquad \{\tilde{u}_1,\tilde{u}_2\}=\{\tilde{p}_{u_1},\tilde{p}_{{u}_2}\}=0\,,
\end{array}
\]
respectively.  These canonical transformations preserve the form of integrals of motion  (\ref{int-ves},\ref{ves-ham-u}), i.e.
they are auto B\"{a}cklund transformations for the Veselova system.
\end{prop}
The proof is a straightforward calculation.

Remind,  that canonical transformation  $(u,p_u)\to(\tilde{u},\tilde{p}_u)$ of valency $c$  determines the Jacobi matrix
\[
V=\left(
    \begin{array}{cc}
      \dfrac{\partial \tilde{u}}{\partial u} &  \dfrac{\partial \tilde{u}}{\partial p_u}  \\ \\
       \dfrac{\partial \tilde{p}_u}{\partial u}  &  \dfrac{\partial \tilde{p}_u}{\partial p_u}  \\
    \end{array}
  \right)\,,
\]
which is a generalized symplectic matrix of valence $c$
\[
V^\top  \Omega V=c\Omega \,,\qquad \Omega=\left(\begin{array}{cc}
            0 & Id \\
            -Id & 0 \\
          \end{array}
        \right)\,,
\]
see \cite{gant}.

 In the first and second cases integrable map $\mathcal B$ determines exact-time discretization of  continuous Hamiltonian flow depending on two parameters $\lambda_{1,2}$ or one parameter $\lambda_1$, respectively. Because trajectories of the Hamiltonian vector field $X$ and original conformally Hamiltonian vector field $Z=\rho X$ coincide to each other,   the same map $\mathcal B$ can be considered also as discretization of the original nonholonomic Veselova system.  Another discretization of the Veselova system  was obtained in the framework of Lagrange formalism  in \cite{igl08}.

 In the third case $\mathcal B$ is a hidden symmetry of the generic level set of the integrals for the Veselova system which is some counterpart of the usual Noether symmetries. Of course, we can not apply  such auto BT to discretization  because when one iterates such BT one will obtain again original elliptic coordinates up to the factor $2$. Other hidden symmetries  can be obtained fixing values of $\lambda_{1,2}$ at the first and second cases.

 For  $\lambda=0$ and $\mu=\sqrt{f(\lambda)}=0$ we have  canonical transformation $\mathcal B$ (\ref{st-b-ves}) on $T^*\mathbb S^2$ which in term of the original coordinates  have the following form
 \[\begin{array}{l}
\tilde{\g}_i=\sqrt{\dfrac{a_jM_j^2+a_kM_k^2}{a_1M_1^2+a_2M_2^2+a_3 M_3^2}-\dfrac{a_i\g_i^2}{g^2}\,},\qquad i\neq j\neq k
\\ \\
\tilde{M}_i=\dfrac{1}{g\,\tilde{g}\,\tilde{\g}_j\,\tilde{\g}_k}\left(
\g_i\bigl(a_j\g_k\tilde{\g}_j^2M_j+a_k\g_j\tilde{\g}_k^2M_k\bigr)-(a_j\tilde{\g}_j^2+a_k\tilde{\g}_k^2)\g_j\g_kM_i\right)\,.
\end{array}
\]
Here $g=\sqrt{a_1\g_1^2+a_2\g_2^2+a_3\g_3^2}$ and $\tilde{g}=\sqrt{a_1\tilde{\g}_1^2+a_2\tilde{\g}_2^2+a_3\tilde{\g}_3^2}$ are the last Jacobi multipliers related to each other via integrals of motion:
\[
g\cdot\tilde{g}=\sqrt{\dfrac{a_1a_2a_3H_2}{H_1}}\,.
\]
It is easy to prove that this transformation  $\mathcal B: (\g,M)\to(\tilde{\g},\tilde{M}) $ preserve the form of the Poisson bivector $P$ (\ref{poi-ves}) and the form of integrals of motion, i.e.
 \[C_1=(\g,\g)=(\tilde{\g},\tilde{\g})=1\,,\qquad C_2=(\g,AM)=(\tilde{\g},A\tilde{M})=0\,,\]
 and
\[
H_1=(M,AM)=(\tilde{M},A\tilde{M}),\qquad
H_2=(M,M)-(\g,M)^2=(\tilde{M},\tilde{M})-(\tilde{\g},\tilde{M})^2\,.
\]
Below we use this hidden symmetry of the level set manifold to construction of new integrable systems on $T^*\mathbb S^2$.

\section{Hetero B\"{a}cklund transformations and new integrable systems on the sphere}
\setcounter{equation}{0}
  The counterpart of the  hetero  B\"{a}cklund  transformations  for finite dimensional  integrable systems  has to be a canonical transformation,  which has to relate  two different systems of the Hamilton-Jacobi equations
   \bq \label{Eq-HJt-2}
   H_i\left(u,\dfrac{\partial S}{\partial u}\right)=h_i\qquad\mbox{and}\qquad \tilde{H}_i\left(\tilde{u},\dfrac{\partial \tilde{S}}{\partial \tilde{u}}\right)=\tilde{h}_i
\eq
and has to satisfy some additional conditions which allow to get a non-trivial, usable  and efficient theory. In \cite{ts15a,ts15b,ts15c}
 we postulated that  $H_i$ are {simultaneously separable}  in $u$ and $\tilde{u}$ variables. This allows us to apply  the standard Jacobi method to construction of $\tilde{H}_i$:
 \begin{enumerate}
  \item take Hamilton-Jacobi equation $H=E$ separable in variables $u,p_u$;
  \item make auto B\"{a}cklund transformation of variables $(u,p_u)\to(\tilde{u},\tilde{p}_u)$, which conserves not only the Hamiltonian character of the equations of motion, but also the form of  Hamilton-Jacobi equation;
  \item substitute new canonical variables $\tilde{u},\tilde{p}_{u}$ into the suitable separated relations  and  obtain new  integrable systems.
\end{enumerate}
Step two is a simple technical exercise in the framework of the Abel theory  due to its various implementations in modern cryptography  \cite{cant87}.

\subsection{First  bi-Hamiltonian system}
Integrals of motion (\ref{ves-ham-u}) satisfy separated relations
\bq\label{sep-rel-u}
\varphi(u_k)p_{u_k}^2=H_1u_k-H_2\,,\qquad k=1,2\,,
\eq
and similar in new variables
\bq\label{sep-rel-ut}
\varphi(\tilde{u}_k)\tilde{p}_{u_k}^2=H_1\tilde{u}_k-H_2\,,\qquad k=1,2\,,
\eq
i.e. $H_{1,2}$ are {simultaneously separable}  in $u$ and $\tilde{u}$-variables.
If we substitute new canonical variables $\tilde{u}_{1,2}$ and $\tilde{p}_{u_{1,2}}$ on $T^*\mathbb S^2$ into the following separation relations
\bq\label{sep-rel-new1}
2\upsilon_1\equiv\varphi(\tilde{u}_1)\cdot\tilde{p}_{u_1}^2=\tilde{H}_1+\tilde{H}_2\,,\qquad 2\upsilon_2\equiv\varphi(\tilde{u}_2)\cdot\tilde{p}_{u_2}^2=\tilde{H}_1-\tilde{H}_2
\eq
and solve the resulting equations with respect to $\tilde{H}_{1,2}$
\[
\tilde{H}_1=\upsilon_1+\upsilon_2=\dfrac{(\tilde{u}_1+\tilde{u}_2)H_1}{2}-H_2\,,\qquad \tilde{H}_2=\upsilon_1-\upsilon_2=\dfrac{(\tilde{u}_1-\tilde{u}_2)H_1}{2}\,.
\]
one gets some additive deformation $\tilde{H_1}=-H_2+\Delta H$ of the second integral of motion $H_2$ in (\ref{ves-ham-u}).
\begin{rem}
These integrals $\tilde{H}_{1,2}$ can be obtained using  definitions (\ref{g2-x23}) of  the symmetric functions on $x_{3,4}=\tilde{u}_{1,2}$ without calculations of much more complicated  expressions for $\tilde{u}_{1,2}$ and $\tilde{p}_{u_{1,2}}$.
\end{rem}

These new Hamiltonians $\tilde{H}_{1,2}$ are in involution with respect to compatible Poisson brackets
\[
\{\tilde{u}_i,\tilde{p}_{u_j}\}=\delta_{i,j}\,,\qquad \{\tilde{u}_1,\tilde{u}_2\}=\{\tilde{p}_{u_1},\tilde{p}_{u_2}\}=0\,,
\]
 and
\[
\{\tilde{u}_i,\tilde{p}_{u_j}\}'=\lambda_i^{-1}\delta_{i,j}\,,\qquad \{\tilde{u}_1,\tilde{u}_2\}'=\{\tilde{p}_{u_1},\tilde{p}_{u_2}\}'=0\,.
\]
Using the corresponding bivectors $P$ and $P'$ it is easy to prove that vector field
\[X=Pd(\upsilon_1+\upsilon_2)=P'd\left(\dfrac{\upsilon_1^2+\upsilon_2^2}{2}\right)
\]
is bi-Hamiltonian vector field. This trivial in $\tilde{u}_{1,2}$ and $\tilde{p}_{u_{1,2}}$ variables Hamiltonian $\tilde{H}=\tilde{H}_1$ has more complicated form in original  variables:
\bq\label{tH-u}
\tilde{H}=\upsilon_1+\upsilon_2=\dfrac{1}{g^2(u_1-u_2)^2}\left(\eta_1\varphi(u_1)p_{u_1}^2+\dfrac{\varphi(u_1)\varphi(u_2)\,p_{u_1}p_{u_2}}{2}+\eta_2\varphi(u_2)p_{u_2}^2\right)\,,
\eq
where $\varphi(u)$ is given by (\ref{ves-phi}),
\[\begin{array}{l}
\eta_1=u_2+u_1u_2\bigl(u_2\alpha_2-\alpha_1-(2u_2-u_1)u_2\alpha_3\bigr)\,,\\
\\
\eta_2=u_1+u_1u_2\bigl(u_1\alpha_2-\alpha_1-(2u_1-u_2)u_1\alpha_3)
\end{array}
\]
and
\[
\alpha_1=a_1+a_2+a_3\,,\qquad \alpha_2=a_1a_2+a_1a_3+a_2a_3\,,\qquad \alpha_3=a_1a_2a_3\,.
\]
Second integral of motion $\tilde{H}_2=\upsilon_1-\upsilon_2$ is an algebraic function and, therefore, we present another function on $\lambda_{1,2}$ which  is the polynomial of fourth order in momenta
\bq\label{tK-u}
\tilde{K}=4\upsilon_1\upsilon_2=\left(\dfrac{u_1p_{u_1}-u_2p_{u_2}}{u_1-u_2}\right)^2\cdot \dfrac{\varphi(u_1)\varphi(u_2)}{u_1u_2(u_1-u_2)}\left(\dfrac{\varphi(u_1)p_{u_1}^2}{u_1}-\dfrac{\varphi(u_2)p_{u_2}^2}{u_2}\right)\,.
\eq
Integral of motion $\tilde{K}$ is factored on two polynomials of second order in momenta which do not commute with $\tilde{H}$. In three dimensional case  similar examples of quartic integrals of motion are discussed in \cite{ts15}.

In redundant  variables $\g$ and $M$ these Hamiltonians have the following form:
\bq\label{ves-new1}
\tilde{H}=g^{-2}(M,\Omega)-(M,M)+2(\g,M)^2\,,\qquad \tilde{K}=4a_1a_2a_3(\g,M)^2\cdot\Bigl((M,M)-(\g,M)^2\Bigr)\,.
\eq
\begin{prop}
If $(\g,\g)=1$ and $(\g,\Omega)=0$,  Hamiltonians (\ref{ves-new1}) are in involution
\[\{\tilde{H},\tilde{K}\}=0\]
with respect to the Poisson bracket defined by the Poisson bivector $P$ (\ref{poi-ves}).
\end{prop}
The proof is a straightforward calculation.

Using more complicated separation relations instead of (\ref{sep-rel-new1}) we can get more complicated Hamiltonians on $T^*\mathbb S^2$ which have a natural form $\tilde{H}=T+V$ and  an additional quartic integral of motion \cite{ts15a,ts15b,ts15c,ts16m}.

\subsection{Second bi-Hamiltonian system}
Let us rewrite separated relations (\ref{sep-rel-ut}) for original integrals of motion (\ref{ves-ham-u}) in the following form
\[
\dfrac{\varphi(\tilde{u}_k)\tilde{p}_{u_k}^2}{\tilde{u}_k}=H_1-\dfrac{H_2}{u_k}\,,\qquad k=1,2\,.
\]
If we substitute new canonical variables $\tilde{u}_{1,2}$ and $\tilde{p}_{u_{1,2}}$ on $T^*\mathbb S^2$ into the following separation relations
\bq\label{sep-rel-new2}
2\nu_1\equiv\dfrac{\varphi(\tilde{u}_1)\tilde{p}_{u_1}^2}{u_1}=\hat{H}_1+\hat{H}_2\,,\qquad 2\nu_2\equiv\dfrac{\varphi(\tilde{u}_2)\tilde{p}_{u_2}^2}{u_2}=\hat{H}_1-\hat{H}_2
\eq
and solve the resulting equations with respect to $\hat{H}_{1,2}$
\[
\hat{H}_1=\nu_1+\nu_2=H_1-\dfrac{(\tilde{u}_1+\tilde{u}_2)H_2}{2\tilde{u}_1\tilde{u}_2}\,,\qquad
\tilde{H}_2=\nu_1-\nu_2=\dfrac{(\tilde{u}_1-\tilde{u}_2)H_2}{2\tilde{u}_1\tilde{u}_2}
\]
one gets some additive deformation $\hat{H}_1=H_1-\Delta H_1$ of the first integral of motion $H_1$ in (\ref{ves-ham-u}).

These new Hamiltonians $\hat{H}_{1,2}$  are in involution with respect to compatible Poisson brackets
\[
\{\tilde{u}_i,\tilde{p}_{u_j}\}=\delta_{i,j}\,,\qquad \{\tilde{u}_1,\tilde{u}_2\}=\{\tilde{p}_{u_1},\tilde{p}_{u_2}\}=0\,,
\]
 and
\[
\{\tilde{u}_i,\tilde{p}_{u_j}\}''=\nu_i^{-1}\delta_{i,j}\,,\qquad \{\tilde{u}_1,\tilde{u}_2\}''=\{\tilde{p}_{u_1},\tilde{p}_{u_2}\}''=0\,.
\]
Using the corresponding bivectors $P$ and $P''$ it is easy to prove that vector field
\[X=Pd(\nu_1+\nu_2)=P''d\left(\dfrac{\nu_1^2+\nu_2^2}{2}\right)
\]
is bi-Hamiltonian vector field. This trivial in $\tilde{u}_{1,2}$ and $\tilde{p}_{u_{1,2}}$ variables Hamiltonian $\hat{H}=\hat{H}_1$ has more complicated form in original  variables:
\bq\label{hH-u}
\hat{H}=\nu_1+\nu_2=\dfrac{1}{(u_1-u_2)^2}\left(
\zeta_1\varphi(u_1)p_{u_1}^2+\dfrac{\varphi(u_1)\varphi(u_2)p_{u_1}p_{u_2}}{2}+\zeta_2\varphi(u_2)p_{u_2}^2
\right)\,,
\eq
where $\varphi(u)$ is given by (\ref{ves-phi}),
\[\begin{array}{l}
\zeta_1=2u_1-u_2+u_1u_2(\alpha_2u_2-\alpha_1)-\alpha_3u_1^2u_2^2\,,\\
\\
\zeta_2=2u_2-u_1+u_1u_2(\alpha_2u_1-\alpha_1)-\alpha_3u_1^2u_2^2\,.
\end{array}
\]
Second integral of motion $\hat{H}_2=\nu_1-\nu_2$ is an algebraic function and, therefore, we present another function on $\nu_{1,2}$ which  is the polynomial of fourth order in momenta
\bq\label{hK-u}
\hat{K}=4\nu_1\nu_2=\left(\dfrac{u_1p_{u_1}-u_2p_{u_2}}{u_1-u_2}\right)^2\cdot \dfrac{\varphi(u_1)\varphi(u_2)}{u_1u_2(u_1-u_2)}\left(\varphi(u_1)p_{u_1}^2-\varphi(u_2)p_{u_2}^2\right)\,.
\eq
Integral of motion $\hat{K}$ is factored on two polynomials of second order in momenta and the first factor coincides with the first factor in $\tilde{K}$ (\ref{tK-u}).

In redundant  variables $\g$ and $M$ these Hamiltonians have the following form:
\bq\label{ves-new2}
\hat{H}=(M,\Omega)-g^2(M,M)=(M,\hat{\Omega})\,,\qquad \hat{K}=4g^2(\g,M)^2(M,\Omega)\,,
\eq
where $\hat{\Omega}=\bigl(A-(\g,A\g)\bigr)M$ is  the vector of  "angular velocity" depending on $\g$ variables.
\begin{prop}
If $(\g,\g)=1$ and $(\g,\Omega)=0$,  Hamiltonians (\ref{ves-new2}) are in involution
\[\{\hat{H},\hat{K}\}=0\]
with respect to the Poisson bracket defined by the Poisson bivector $P$ (\ref{poi-ves}).
\end{prop}
The proof is a straightforward calculation.

The corresponding conformally Hamiltonian vector field
\[
\hat{Z}=\dfrac{1}{2g}\,Pd\hat{H}=\dfrac{1}{2g}\,Pd\Bigl(H+\Delta H\Bigr)=Z+\Delta{Z}
\]
is additive deformation of the original vector field $Z$ (\ref{ves-eqm}) for the Veselova system:
\bq\label{ves-eqm-new}
\begin{array}{l}
 \dfrac{d}{d\tau} {\gamma}=\gamma\times \hat{\Omega}-(\g,M)\cdot\g\times A\g\,,\\
 \\
 \dfrac{d}{d\tau}{M}=M\times \hat{\Omega} +\lambda \gamma+\dfrac{1}{2}\cdot \g\times \dfrac{\partial \hat{H}}{\partial \g}
 +\Bigl(M-(\g,M)\g\Bigr)\times AM\,.
 \end{array}
\eq
Such nonstandard equations of motion  may appear in the study of a wide range of fields such  as  control theory,  seismology, biological  systems, in the study of a self graviting stellar gas cloud, optoelectronics, fluid mechanics etc.
 \begin{rem}
 By adding to the Hamiltonian $H_1$ (\ref{int-ves}) one of the integrable potentials  $V(\g)$ from \cite{dr98,fed04} we change equation of motion
 \[
 \dfrac{d}{d\tau} {\gamma}=\gamma\times \Omega\,,\quad \dfrac{d}{d\tau}{M}=M\times \Omega +\dfrac{1}{2}\cdot \g\times \dfrac{\partial V}{\partial \g}\,,
\]
  and obtain new canonical variables on $T^*\mathbb S^2$ after suitable B\"{a}cklund transformation. For instance, if
$V=b(I_1x_1^2+I_2x_2^2+I_3x_3^2)$ \cite{ves86},    we have to  change original separation relations (\ref{sep-rel-u}) on
 \[
 \varphi(u_k)p_{u_k}^2=bu_k^2+u_kH'_1-H'_2\,,\qquad b\in\mathbb R\,.
 \]
 After B\"{a}cklund transformation for this system one gets new canonical variables on $T^*\mathbb S^2$ depending on parameter $b$.
 Substituting these variables into the separation relations (\ref{sep-rel-new2}) we obtain Hamiltonian
 \bq\label{def-pot1}
 \hat{H}'=\hat{H}+\dfrac{4\sqrt{b}\sqrt{a_1a_2a_3}}{u_1-u_2}\left(u_2^2\varphi(u_1)p_{u_1}-u_1^2\varphi(u_2)p_{u_2}\right)
 +b(\alpha_3u_1^2+u_2^2-\alpha_1u_1u_2+2u_1+2u_2)\,,
 \eq
 with potential depending on velocities.  In this case second integrals of motion is the polynomial of fourth order in momenta which does not factorized on two polynomials of second order, see other similar examples in \cite{ts15b,ts15c}.
 \end{rem}

\subsection{Bi-Hamiltonian systems on the two-dimensional sphere}
A special class of natural Hamiltonian systems are geodesic flows, i.e., natural Hamiltonian systems with zero potential. According to the  Maupertuis principle, an integrable natural Hamiltonian system immediately gives a family of integrable geodesic flows. If the integral of the system is polynomial in momenta, the integrals of the geodesic flows are also polynomial of the same degree. There are a few examples of integrable geodesic flows on the sphere $\mathbb S^2$ whose integrals are polynomials in momenta of degree four, see \cite{bol95,bj04,kiy01,val13,yeh13} and references within.

Using the following anzats for integrals of motion on $T^*\mathbb S^2$
\[
H=\frac{f_1(u_1, u_2)p_{u_1}^2+f_2(u_1, u_2)p_{u_1}p_{u_2}+f_3(u_1, u_2)p_{u_2}^2}{(u_1-u_2)^2}\,,
\]
\[
K=\left(\dfrac{u_1p_{u_1}-u_2p_{u_2}}{u_1-u_2}\right)^2\Bigl(h_1(u_1,u_2)p_{u_1}^2+h_2(u_1,u_2)p_{u_2}^2\Bigr)\,,
\]
where $f_i$ and $h_k$ are arbitrary functions on elliptic coordinates $u_{1,2}$ on the sphere, we can prove that equation $\{H,K\}=0$ has two complete solutions  (\ref{tH-u}-\ref{tK-u}) and (\ref{hH-u}-\ref{hK-u}). There is also one partial solution at $a_1=a_2$ associated with the Kovalevskaya systems in the dynamics of a rigid body.

Let us consider the two dimensional unit sphere $\mathbb S^2$  as an imbedded manifold in $\mathbb R^3$
 \[
 \mathbb S^2=\{x\in \mathbb R^3;\quad (x,x)=1\}\,.
 \]
Hence its cotangent bundle $T^*\mathbb S^2$ can be realized as a subvariety of $T^*\mathbb R^3$
  \[
 T^*\mathbb S^2=\{(x,p)\in T^*\mathbb R^3;\quad (x,x)=1,\quad(p,x)=0\}\,.
 \]
 Here $x=(x_1,x_2,x_3)$ and $p=(p_1,p_2,p_3)$ are canonical coordinates in $T^*\mathbb R^3$.

 We also will use coordinates  $x=(x_1,x_2,x_3)$ and  $J=(J_1,J_2,J_3)$  on the Euclidean algebra $e(3)^*$ with the Lie-Poisson
brackets
\begin{equation}\label{e3}
\,\qquad \bigl\{J_i\,,J_j\,\bigr\}=\varepsilon_{ijk}J_k\,, \qquad
\bigl\{J_i\,,x_j\,\bigr\}=\varepsilon_{ijk}x_k \,, \qquad
\bigl\{x_i\,,x_j\,\bigr\}=0\,,
\end{equation}
where $\varepsilon_{ijk}$ is the totally skew-symmetric tensor. Fixing  values
\[ (x,x)=1,\qquad (x,J)=0\]
 of the Casimir functions one gets symplectic leaf of $e^*(3)$ which is a four-dimensional symplectic manifold
 equivalent to $T^*\mathbb S^2$ \cite{bol95}.

\begin{rem}
Variables  $(x,J)$  are related with the original Veselova variables by the following transformation \cite{bm05}:
\[
x=g^{-1}I^{-1/2}\g\,,\qquad J=gI^{1/2}\Omega\,,
\]
which reduce Poisson bivector (\ref{poi-ves}) to the standard Poisson  bivector associated to (\ref{e3})
 \[
P=\left(\begin{matrix}
        0 &  0 &  0 & 0  & x_3  & -x_2 \\
       0 &  0 &   0 &  -x_3  & 0   & x_1\\
       0 &  0 &  0 &  x_2  &-x_1& 0 \\
     0 & x_3 & -x_2 &   0 & J_3 &-J_2  \\
       -x_3 & 0  & x_1 &-J_3 &   0 & J_1\\
      x_2 & -x_1 &0  &J_2  &-J_1  &   0 \\
    \end{matrix}\right)
  \,,
\]
up to the factor $\sqrt{a_1a_2a_3}$.
\end{rem}

Let us rewrite Hamiltonians  (\ref{tH-u}-\ref{tK-u}) and (\ref{hH-u}-\ref{hK-u}) in term of redundand variables $(x,p)$ and $(x,J)$ on $T^*\mathbb R^3$.
\begin{prop}
If $A$ is an arbitrary symmetric matrix defining two functions on $T^*\mathbb R^3$
\ben
\hat{H}&=&(x,x)(J,AJ)-(x,Ax)(J,J)=\left|
                                    \begin{array}{cc}
                                      (x,x)   & (x,Ax) \\
                                      (J,J) & (J,AJ) \\
                                    \end{array}
                                  \right|\nn\\
\nn\\
&=&(x\times p,A(x\times p))-(x,Ax)p^2
\en
and
\[
\hat{K}=(x,AJ)^2(J,J)=(Ax,x\times p)p^2=[x,p,Ax]^2p^2\,,
\]
where $[x,p,Ax]^2$ is a Gram determinant
\[
[x,p,Ax]^2=\left|
             \begin{array}{ccc}
               (x,x) & (x,p) & (x,Ax) \\
               (x,p) & (p,p) & (p,Ax) \\
               (x,Ax) & (p,Ax) & (Ax,Ax)\\
             \end{array}
           \right|\,,
\]
then the Poisson brackets  (\ref{e3})  between $\hat{H}$ and $\hat{K}$ is equal to
\[
\{\hat{H},\hat{K}\}=(x,J)(x,AJ)(J,J)\cdot\{(J,AJ),(x,Ax)\}\,.
\]
Hence, $\hat{H}$ and $\hat{K}$ are in involution at $(x,J)=0$, i.e. on the cotangent bundle to the unit sphere $T^*\mathbb S^2$,   and also on the zero level set of the second Hamiltonian.
\end{prop}
The proof is a straightforward calculation.

\begin{prop}
If $A$ is an arbitrary symmetric matrix defining two functions on $T^*\mathbb R^3$
\ben
\tilde{H}&=&(x,Ax)\hat{H}-(x,x)(x,AJ)^2
=\left|
                                    \begin{array}{ccc}
                                      (x,x)   & (x,Ax) &0 \\
                                      (J,J) & (J,AJ) &\sqrt{2}\,(x,AJ)\\
                                      0& \sqrt{2}\,(x,AJ)&(x,Ax)\\
                                    \end{array}
                                  \right|
\en
and
\[
\tilde{K}=(x,AJ)^2\Bigl((x,Ax)(J,AJ)- (x,AJ)^2 \Bigr)=[x,p,Ax]^2\left|
                                                                  \begin{array}{cc}
                                                                    (x,Ax) & (x,AJ) \\
                                                                    (x,AJ) & (J,AJ) \\
                                                                  \end{array}
                                                                \right|\,,
\]
where $=[x,p,Ax]^2$ is a Gram determinant, then the Poisson brackets  (\ref{e3})  between $\tilde{H}$ and $\tilde{K}$ is equal to
\[
\{\tilde{H},\tilde{K}\}=(x,J)(x,Ax)^2(x,AJ)\cdot\Bigl((J,AJ)\{(J,AJ),(x,Ax)\}+4(x,AJ)\{(J,AJ),(x,AJ)\}\Bigr)\,.
\]
Hence, $\tilde{H}$ and $\tilde{K}$ are in involution at $(x,J)=0$, i.e. on the cotangent bundle to the unit sphere $T^*\mathbb S^2$,   and also on the zero level set of the second Hamiltonian.
\end{prop}
The proof is a straightforward calculation.

In elliptic coordinates on the sphere these Hamiltonians have the form (\ref{hH-u}-\ref{hK-u}) and (\ref{tH-u}-\ref{tK-u}) and, therefore, we have two bi-Hamiltonian system on  $T^*\mathbb S^2$. Using natural Hamiltonians  $\hat{H}$ and $\tilde{H}$
 we can get globally defined  integrable metrics and geodesic flows on the sphere $\mathbb S^2$ \cite{bol95, bj04} whose integrals are polynomials in redundant momenta of degree four.  We believe that our bi-Hamiltonian systems does not overlap with the implicit systems discussed in \cite{kiy01,val13,yeh13}.
We can try to add some integrable potentials to the Hamiltonians  $\hat{H}$  and $\tilde{H}$, for instance potential (\ref{def-pot1}), and try to obtain multidimensional generalizations of these bi-Hamiltonian systems. Discussion of these topics to look beyond the main content of this paper.

\section{Conclusion}
We study symmetries of the  integrals of motion associated with the nonholonomic Veselova system using well-known divisor arithmetic on hyperelliptic curve of genus two \cite{cant87}. In addition to the standard one-parametric auto B\"{a}cklund transformations \cite{fed00,kuz02} we also discuss symmetries related to a general divisor doubling \cite{cost12} in order to get canonical transformation of valence two.  Then we use these discrete symmetries in order to get new canonical variables on the phase space and new integrable Hamiltonian and conformally Hamiltonian systems in the framework of the Jacobi method.

 With a pure mathematical point of view  nonholonomic Veselova system is equivalent to the nonholonomic Chaplygin ball \cite{ts12,ts12b}. It allows us to get new integrable deformations of the Chaplygin ball with integral of motion of fourth order in momenta and try to describe the corresponding physical model. We plan to study symmetries of integrals of motion and integrable deformations  of the Chaplygin ball in a forthcoming publication.

The work was supported by the Russian Science Foundation (project  15-11-30007).

\end{document}